\documentclass[rmp,nobibnotes,superscriptaddress,amsmath,amssymb,amsfonts,twocolumn]{revtex4-1}
\usepackage{array,multirow,graphicx}
\usepackage{xcolor}
\usepackage{ulem}
\usepackage[latin1]{inputenc}
\usepackage{caption}
\usepackage{scrextend}
\usepackage{hyperref}
\usepackage{float}
\usepackage{subcaption}
\usepackage{gensymb}

\def\sss{\scriptscriptstyle\rm}
\def\br{{\bf r}}
\def\bq{{\bf q}}

\def\xc{_{\sss XC}}
\def\bn{\boldsymbol{\nabla}}
\def\ext{_{\rm ext}}
\def\s{_{\sss S}}
\def\H{_{\sss H}}

\begin{document}

\normalem 

\title{Control of magnons via ultrafast magnetization modulation}
\author{N. Singh}
\affiliation{Max-Born-Institut  f\"ur Nichtlineare Optik und Kurzzeitspektroskopie, Max-Born-Strasse 2A, 12489 Berlin, Germany.}
\author{P. Elliott}
\affiliation{Max-Born-Institut  f\"ur Nichtlineare Optik und Kurzzeitspektroskopie, Max-Born-Strasse 2A, 12489 Berlin, Germany.}
\author{J.K. Dewhurst}
\affiliation{Max-Planck-Institut f\"ur Mikrostrukturphysik, Weinberg 2, D-06120 Halle, Germany.}
\author{S. Sharma}
\affiliation{Max-Born-Institut  f\"ur Nichtlineare Optik und Kurzzeitspektroskopie, Max-Born-Strasse 2A, 12489 Berlin, Germany.}

\date{\today}

\begin{abstract}
We demonstrate optical control of magnons using femtosecond laser pulses by performing \emph{ab-initio} real-time time-dependent density functional theory (TDDFT) simulations. We predict that the spin-wave dynamics in Fe$_{50}$Ni$_{50}$ can be manipulated by tailoring the applied laser pulse via three distinct mechanisms: (1) element selective destruction of magnon modes depending on the laser intensity, (2) delay dependent freezing of the magnon mode into a transient non-collinear state (where delay is in the pulse peak with respect to the start of simulations), and (3) optically induced inter sublattice transfer (OISTR) driven renormalization of the optical magnon frequency. Harnessing such processes would significantly speed up magnonic devices.
\end{abstract}

\maketitle


Ultrashort laser technology has emerged as an unexpected, but extremely promising, tool for manipulating the magnetic properties of materials on femtosecond (fs) timescales. The laser light excites the material's constituent electrons into a non-equilibrium state, which then alters the magnetic properties. This is due to the fact that, fundamentally, magnetism arises due to the angular momentum of the electrons, especially the intrinsic angular momentum or spin-moment of the electrons. Thus, the interaction between light and electrons may be used to influence the spin dynamics and hence magnetic properties of materials. If such processes can be successfully exploited, they may lead to electronic devices operating at speeds several orders of magnitude faster than those currently available. 

The last few decades have seen the emergence of the separate field of spintronics, where manipulating the spin of the electron is established as fundamental to realizing the next technological leap forward in terms of speed, size, and energy efficiency. One physical phenomena studied in spintronics are the collective, low energy (meV), spin-wave quasi-particle excitations known as magnons. These spin-waves have several favourable properties\cite{QZES15}, such as carrying spin-currents without suffering from the limitations encountered by charge currents in conventional electronics. However it is not known how magnons react to ultrafast changes in the magnetization due to applied laser pulses. Thus, the goal of this work is to investigate how magnon modes response to ultrafast laser pulses, thus extending magnonics into the fs regime. 

As the system is strongly excited by the pump laser, a theoretical method valid in the non-linear regime is required in order to simulate such a situation.  However, the usual theoretical approach to study magnons: atomistic spin dynamics using the Heisenberg model \cite{Heisenberg1928} combined with the Landau-Lifshitz-Gilbert equation of motion\cite{MBO18,HACNCON15,EFCOEC14}, is not applicable in this situation. This is due to the dependence on parameterized exchange interactions which assume that the system is in its ground-state and the applied perturbation is very small. In such a situation the exchange parameters can be extracted from \emph{ab-initio} DFT calculations. When the system is pumped into a non-equilibrium state, this is no longer the case and another method must be used. In this work we will use the \emph{ab-initio} method of time-dependent density functional theory (TDDFT).

TDDFT is a formally exact method for treating the dynamics of many-electron systems under the influence of external perturbations\cite{RG84,PE09,SDG14}. TDDFT has recently been shown to successfully predict many processes in spin dynamics, in particular Optical Inter-sublattice Spin Transfer (OISTR) was predicted in Refs. \onlinecite{EMDS16,DESGS18} and then demonstrated experimentally across a wide range of materials and geometries, such as bulk Heusler compounds \cite{S19}, Co/Cu interfaces \cite{CBEMEGDS19}, Ni/Pt multilayers \cite{SGOD18}, and Co/Pt alloys\cite{WKSS20}. For magnon physics, TDDFT has previously been used in the linear regime to predict magnon frequencies and lifetimes\cite{SENDS19,BES11,S98}. It was recently extended to treat magnons in real-time by the authors\cite{SEDGS20} among others\cite{TER20}, which allows the present work to go a step further and observe the effect of non-linear perturbation on the magnon modes. 

While a number of key mechanisms for controlling spin dynamics using ultrafast laser pulses have been identified, e.g. ultrafast demagnetization \cite{BMDB96} whereby loss of spin (or magnetic moment) occurs in less than 100 fs when acted upon by an optical laser pulse or all-optical switching\cite{SHKKTIR07,RVS11} in which the spins switch by 180$\degree$ when excited by the laser, for this work we will focus on the, previously mentioned, OISTR process. The defining signature of OISTR is the transfer of spin from one atom to another, thus modulating the local magnetic moment and disrupting the magnetic interactions between the electrons, both of which may alter the magnon spin-wave states. As these excitations are directly induced by the pump laser, they take place on the timescale of the laser, typically a few femtoseconds. Therefore, the question to be answered is whether this process can be used to control magnon dynamics on fs timescales.

Below we will show three examples of how OISTR can be used to control magnon modes. We demonstrate (1) element selective destruction of magnon modes in multi-component magnetic materials, (2) element selective canting of the magnetic moment i.e. a laser induced transient non-collinear state, and (3) frequency change of selected magnon modes. The system we choose to demonstrate these is  the frequently studied Fe$_{50}$Ni$_{50}$ alloy. This is a multi-sublattice ferro-magnet and is known\cite{MLCP12,RSER15} to have element specific magnetization dynamics making it a good candidate for the present study.

 \begin{figure}[t]
\begin{tabular}{c c}
   \includegraphics[width=0.22\textwidth]{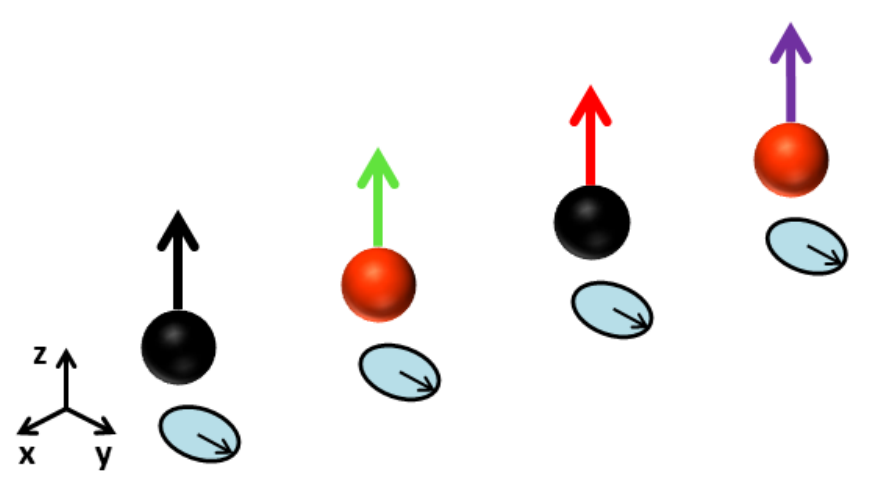} &
    \includegraphics[width=0.22\textwidth]{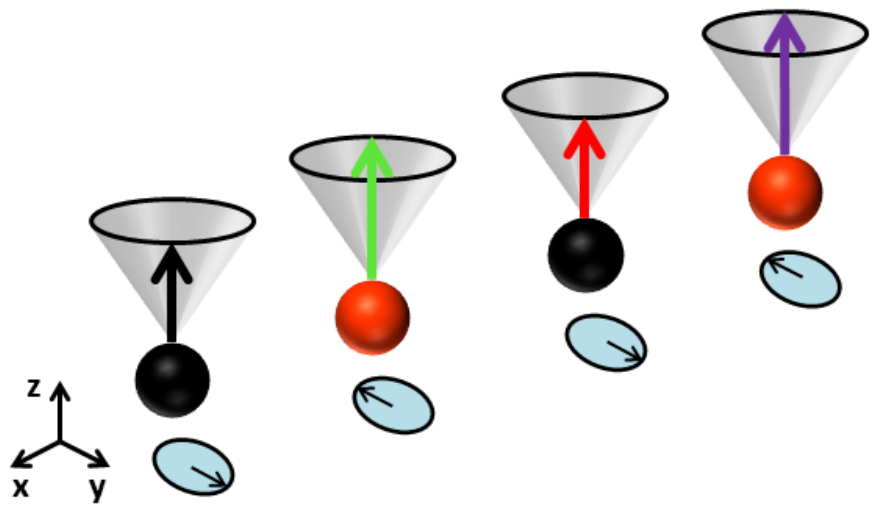} \\
   \small (a)  &  (b) \\
   \includegraphics[width=0.22\textwidth]{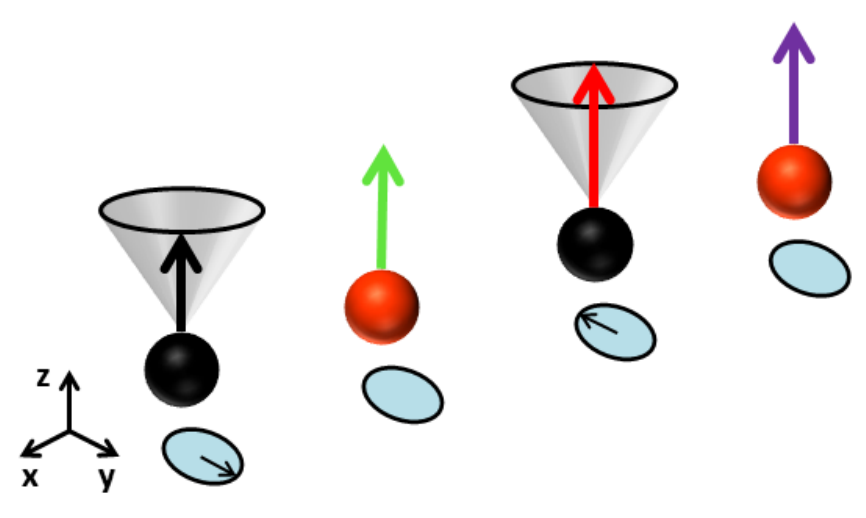} &
    \includegraphics[width=0.22\textwidth]{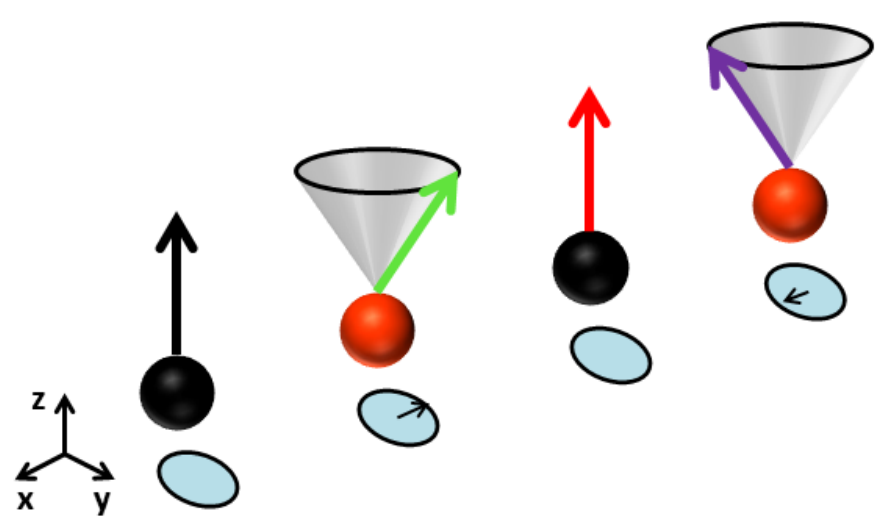} \\
   \small (c)  &  (d) \\
\end{tabular}
\caption{\footnotesize{The four observed modes in a $4$ atom supercell of ferromagnetic alloy, Fe$_{50}$Ni$_{50}$. (a) Goldstone mode, (b) Optical mode, (c) Pure iron mode, and (d) Pure nickel mode}}.
\label{f:modes}
\end{figure}

\begin{figure}[htb]
\includegraphics[width=0.35\textwidth,angle=-90]{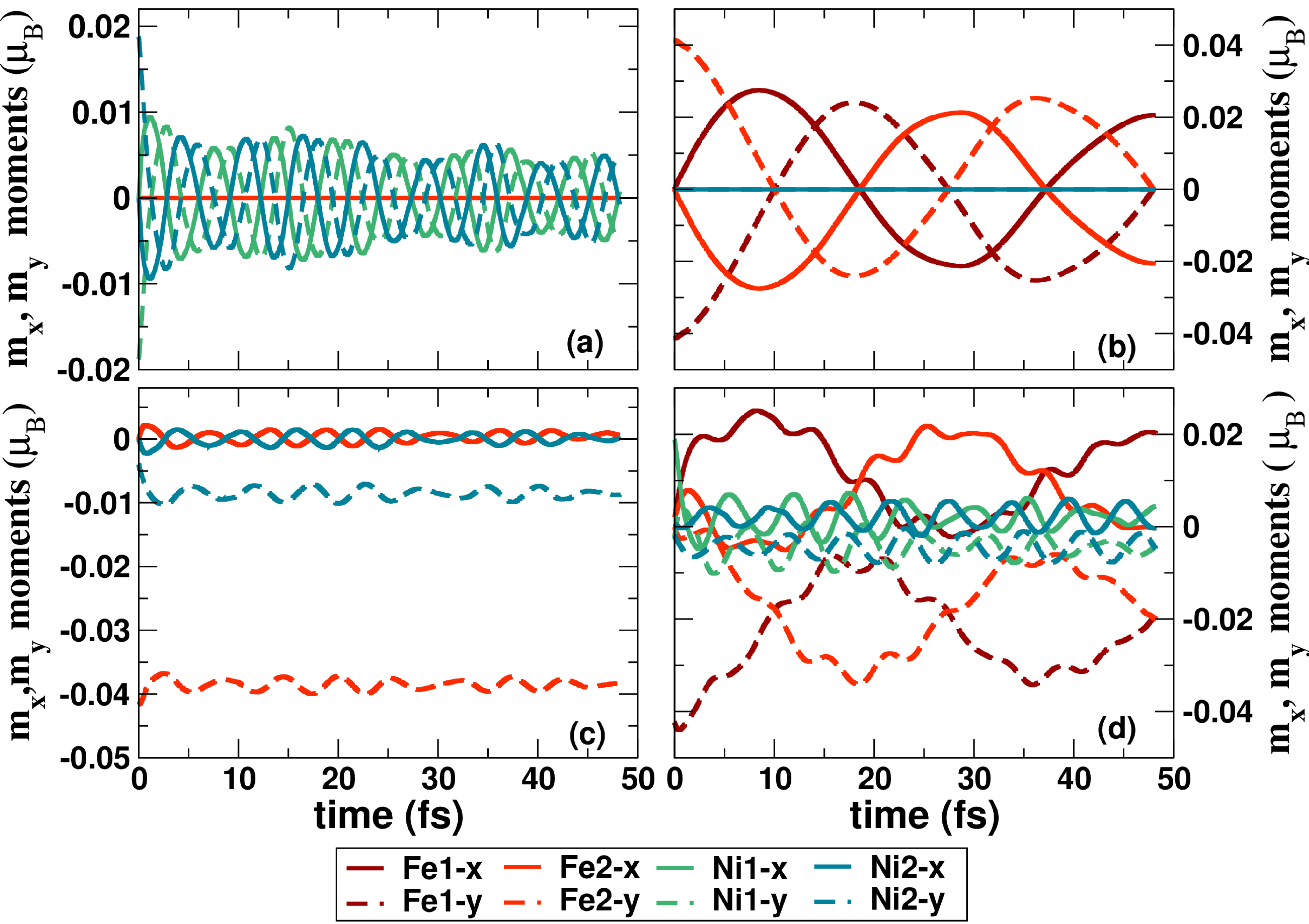} 
\caption{\footnotesize{Oscillation of the transverse (x,y) magnetic moments of the individual nickel and iron atoms in a 4-atom
supercell of Fe$_{50}$Ni$_{50}$ for different initial states. These magnons correspond to momenta q = $\Gamma$, $\pm$ 1/2 $\Gamma X$, and X. Decoupled, element specific magnon modes can be seen for (a) nickel and (b) iron . Coupled Goldstone and optical modes can be seen in (c) and all 4 modes are excited in (d)}.}
\label{f:feni_modes}
\end{figure}


\section*{Results}

To understand the effect on the Fe$_{50}$Ni$_{50}$ magnons of pumping the electrons to an non-equilibrium state using an ultrashort laser pulse, we must first study the unperturbed magnon modes. These are shown in Fig. \ref{f:modes} for a $4$ atom supercell of Fe$_{50}$Ni$_{50}$. The size of the supercell determines how many magnon modes are present in our calculations, however, it is also constrained by the computational power available. With a 4 atom supercell, we sample the magnon wavevectors over the entire first Brillouin zone: the four modes correspond to wave-vectors $\boldsymbol{\Gamma}$, $\pm 1/2 \boldsymbol{\Gamma}\mathbf{X}$, and $\mathbf{X}$ where $\mathbf{X}=(0,0,2\pi/a)$ in cartesian coordinates of fcc primitive unit cell. This wavevector determines the phase difference between adjacent atomic sites, as can be seen in the real-time TDDFT data presented in Fig. \ref{f:feni_modes}.

By choosing an appropriate initial states, we can control which modes are present in our calculations. For the unperturbated system, this allows us to observe and isolate both coupled and decoupled modes (where only one of the magnetic sub-lattices oscillates): 
1) high energy pure Ni mode with $\omega=710$ meV (Fig. \ref{f:feni_modes} (a)). The energy of this mode is higher than the corresponding mode in bulk Ni ($390$ meV). 2) A low energy pure Fe mode with $\omega=90$ meV (see Fig. \ref{f:feni_modes} (b)), the frequency of this mode is also higher than the corresponding mode in bulk Fe ($65$ meV).  The reason for existence of these decoupled modes is the fact that at wave-vector $\bq=\pm 1/2\boldsymbol{\Gamma}\mathbf{X}$, the effective exchange fields acting on an atom, from nearest-neighbor atoms of the other species, cancel leading to only one of the magnetic sub-lattices to oscillate.
The other two modes, out of the four allowed modes, are the coupled Fe and Ni modes: 3) the Goldstone mode ($\omega=0$) whereby all spins tilt together, as seen in Fig. \ref{f:feni_modes} (c) dotted lines and  4) the optical mode, where the Fe and Ni oscillate $180\degree$ out-of-phase with each other, as can also be seen in Fig. \ref{f:feni_modes} (c) full lines. 
The frequency of this mode is $760$ meV, much higher than the $\bq=\mathbf{X}$ mode in either Fe or Ni. 
All these modes can also be excited at the same time, as in Fig. \ref{f:feni_modes} (d). 



 
 
 

\begin{figure}[t]
\includegraphics[width=0.75\columnwidth]{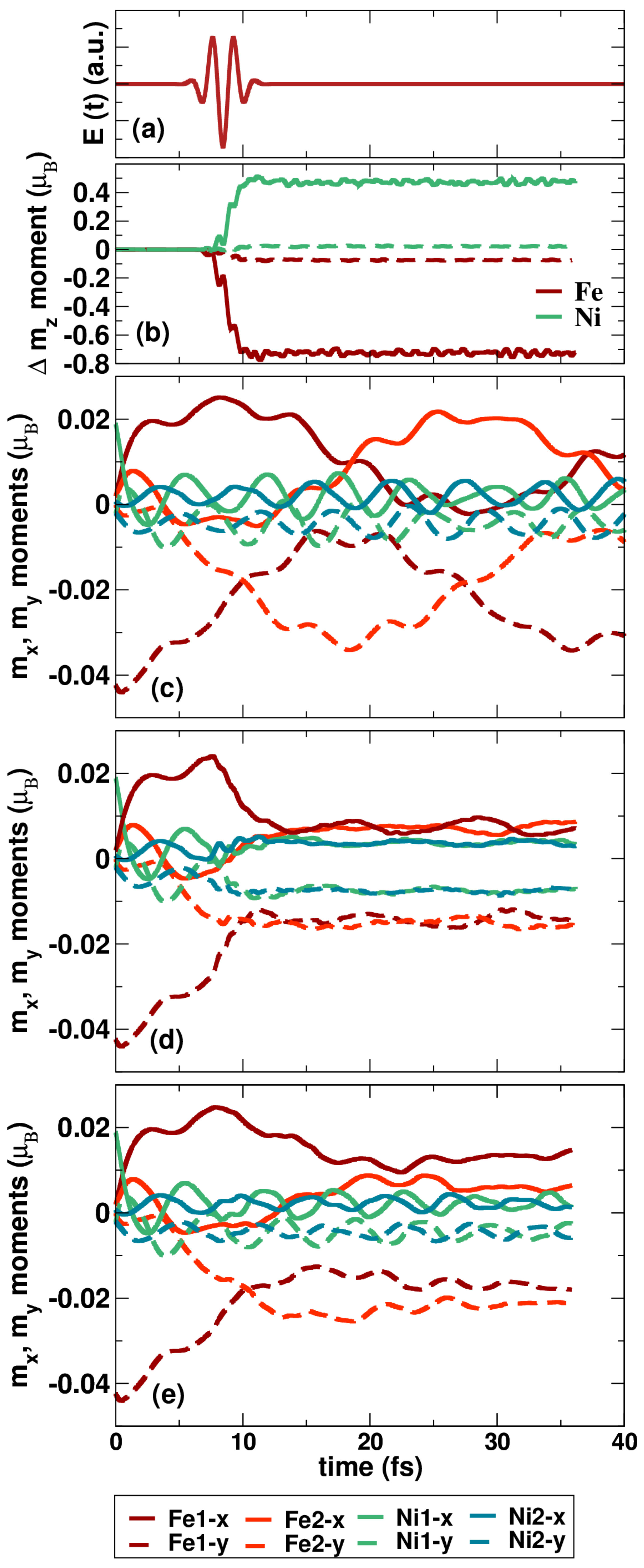} 
\caption{\footnotesize{(a) The electric field profile of the two laser pulses designed to induce OISTR transitions in Fe$_{50}$Ni$_{50}$, both have frequency 2.19eV and FWHM 2.41 fs, but different fluences 9.6807 mJ/cm$^2$ and 0.9537 mJ/cm$^2$. (b) The change in $z$-magnetic moment for iron and nickel with the strong fluence pulse (solid lines) and weak laser pulse (dashed lines). (c) The unperturbed modes. The response of these modes to the strong (d) and weaker (e) laser pulses. In (d) only the Goldstone and optical modes survive, whereas in (e) the pure Fe mode is destroyed while the pure Ni persists.}}.
\label{f:oistr_modes}
\end{figure}


In order to be able to manipulate magnons at ultra-fast time scales we now investigate the behavior of magnon modes under short laser pulses. One of the fastest possible spin response to lasers is via OISTR, primarily driven by minority spin electrons optically excited from one magnetic sub-lattice to another, causing an increase in the moment on the first sub-lattice. In the present work the materials, as well as the laser pulses, are chosen to maximize OISTR: in the Fe$_{50}$Ni$_{50}$ alloy the magnetic moment on the Fe sub-lattice (2.88 $\mu_{\rm B}$) is much higher than on the Ni sub-lattice (0.64 $\mu_{\rm B}$). This causes laser induced optical excitations to transfer minority spin electrons from Ni to Fe, which in turn leads to a increase in the moment on the Ni site, while a corresponding decrease on the Fe site (see Fig. \ref{f:oistr_modes}) (b). The frequency of the laser pulse (2.19 eV) is tuned to optimize this charge transfer. 
Following laser excitation via this OISTR mechanism, the system will relax back to the ground-state on a ps timescale. From our calculations, we see three major effects on magnon dynamics due to OISTR excitations:

\emph{Element selective optical destruction of magnons:} 

The effect of OISTR on the various magnon modes can be  seen in Fig. (\ref{f:oistr_modes})-- strong laser pulse (incident fluence of 9.6807 mJ/cm$^2$ and FWHM of 2.41 fs) effectively destroy both decoupled modes (see Fig. \ref{f:oistr_modes} (d)); the amplitude of the pure Fe magnon mode collapses with only small oscillations remaining which are also quickly damped. Looking at the Ni moments, which initially are a superposition of the pure Ni mode and the optical mode, we see that now only the optical mode exists as the two Ni atoms behave identically (recall that in the pure Ni mode, the two are $180\degree$ out-of-phase). 



These magnon modes show a different dynamics when subjected to a  weaker laser pulse of incident fluence 0.9537 mJ/cm$^2$; the pure nickel mode now survives while the Fe mode is still destroyed (see in Fig. \ref{f:oistr_modes} (e)). In this case the Fe atoms cant with respect to each other with a new, but much reduced, pure Fe mode oscillating about this new configuration. By examining the amount of spin up/down electrons excited on each atom, we find that the Fe atoms have significantly more local optical excitations than Ni. This causes the Fe-Fe exchange coupling to be modified more strongly than the Ni-Ni coupling, explaining the difference in behavior between the two modes. Thus we have found a method by which we can selectively destroy either both Fe and Ni modes or just the Fe mode, on a femtosecond timescale by tuning the fluence of the laser pulse. 
This is an important finding as it not only offers a mechanism of control over magnons but also highlights the fact that the dynamics of element specific magnetization in alloys can greatly differ due to choice of pump pulse\cite{RVS11,MLCP12}.

\begin{figure}[t]
\includegraphics[width=0.205\textwidth,angle=-90]{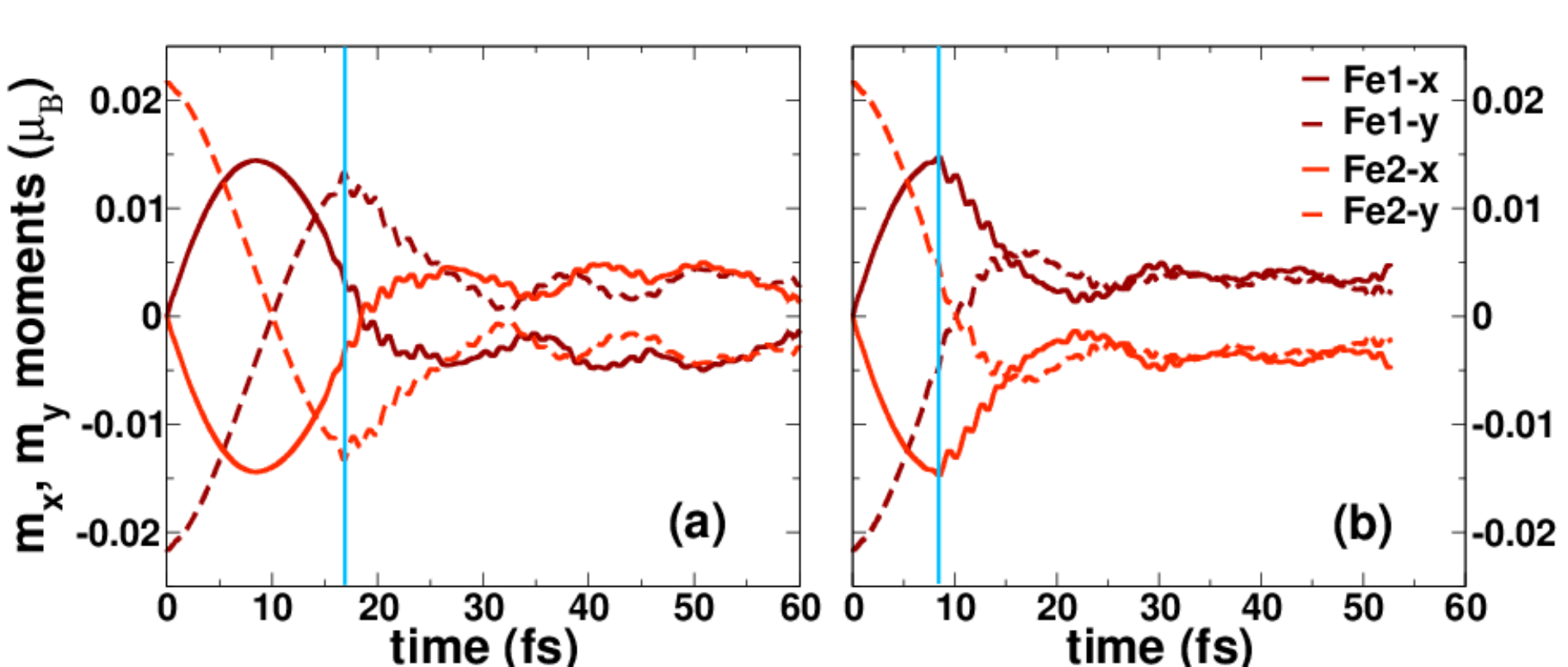} 
\caption{\footnotesize{The canting vector is dependent on the delay time of the laser and is shown relative to the pure Fe mode oscillations. A laser of fluence 0.9537 mJ/cm$^2$ is applied at (a) 16.8 fs and (b) 8.4 fs on the pure iron mode. The vertical lines represents the peak of the laser pulse.}}
\label{f:canting}
\end{figure}

\emph{Control of canting vector:}
Change in relative directions of the moments (i.e. angle between the inter-site spins in a multi-sub-lattice system) in a material can be obtained by rapid destruction of selected magnon modes. We demonstrate, in Fig. \ref{f:canting}, that the canting vector, $\textbf{m}^{\text{Fe1}}(t)-\textbf{m}^{\text{Fe2}}(t)$, can be controlled using the time delay of the laser pulse. 
The effect of the laser pulse is to destroy the magnon mode, but the phase of the magnon mode at the point when the laser is applied determines in which direction the Fe moments eventually point, and thus determines the direction of the canting vector. 
In the first scenario the center of the pulse is chosen to be located at $16.8$ fs when the Fe1y and Fe2y are at their maximum amplitude (see Fig. \ref{f:canting} (a)) and in the second case the center of the pulse is chosen to be at $8.4$ fs which corresponds to the point in time when the Fe1x and Fe2x moments  are at their maximum amplitude (see Fig. \ref{f:canting} (b)). These two time delays in the laser pulses (with respect to the start of simulation) result in different directions of the canting vector. The laser pulse used to obtain this canting is very weak with fluence=0.9537 mJ/cm$^2$. Unlike in the strong laser case of Fig. \ref{f:oistr_modes} (d), where the amplitude of the Fe mode is reduced to zero, in Fig. \ref{f:canting}, the Fe moments remain finite but cease precessing, resulting in a final canted transient state. The main reason behind this is that the laser excitation disrupts the exchange coupling between the nearest Fe atoms, causing the magnon mode to \emph{freeze} into a spin spiral configuration. Extending the delay by half a period of the magnon oscillation will result in a canting vector pointing in opposite direction. This indicates that with a careful choice of laser pulse a ferromagnetic metal can be made to be transiently non-collinear with a certain degree of control over the angle between inter-site spins. 


\begin{figure}[t]
\includegraphics[width=0.35\textwidth,angle=-90]{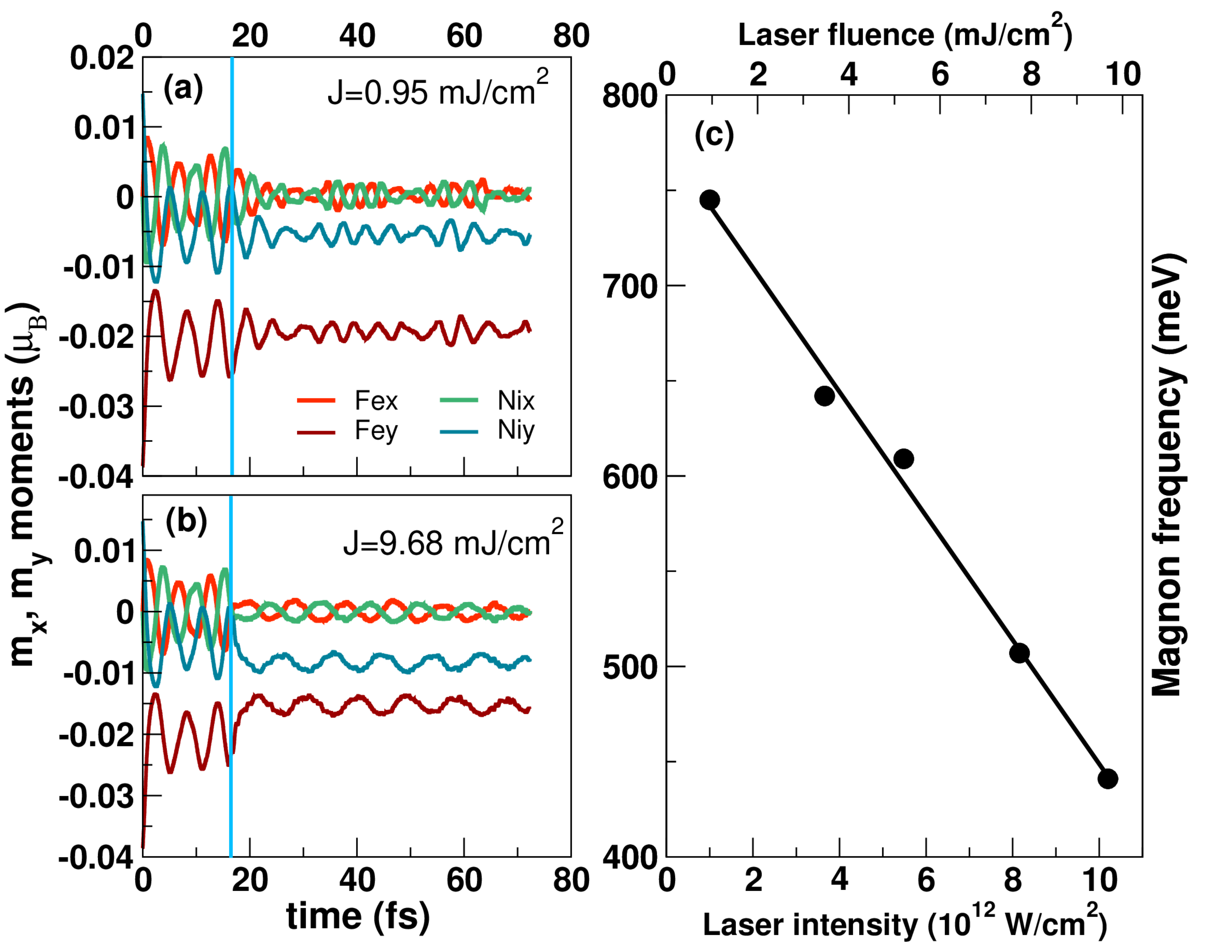} 
\caption{\footnotesize{The change in magnon frequency is shown for two laser pulse of intensities (a) 0.9537 mJ/cm$^2$ and (b) 9.680 mJ/cm$^2$. (c) The magnon frequency decreases with the increase in intensity of applied laser pulse. The vertical line corresponds to the peak of the laser pulse.}}
\label{f:frequency_change}
\end{figure}

\emph{Ultrafast change in magnon frequency:} The frequency of the magnon modes can also be manipulated by pump-laser pulse. To demonstrate this we excite the optical mode and then look at its dynamics under the influence of pump pulses of differing fluences. The results, for two different laser intensities (0.9537 and 9.68 mJ/cm2) are shown in Figs. \ref{f:frequency_change} (a) and (b) where it is clear that the oscillations are strongly influenced by the laser.
Fourier transform of the transverse moment during these oscillations gives the frequency of the magnon mode and this is plotted, as function of laser intensity, in Fig.  \ref{f:frequency_change} (c). The main reason behind this change in frequency is the weakened exchange field between the magnetic sub-lattices \cite{DKJK10,ME14,DSR18} due to two processes, both of which lead to increased screening between the electrons of each atom-- (a) excitation of electrons to excited delocalized states and (b) transfer of localized charge from one atom to the other. 


This implies that the stronger this charge transfer is, the greater the change in the magnon frequency, a fact that is reflected in the linear dependence of the magnon frequency on the pump-pulse fluence (Fig. \ref{f:frequency_change} (c)). At some higher intensity where the charge excitation process saturates, so would the change in the magnon frequency. Thus optical excitations offer  a direct control of frequency of a coupled magnon mode of two sub-lattices via tuning of the fluence of the laser pulse. Since OISTR effects are very strong on AFM coupled systems, we expect very large changes in magnon modes when pumped with lasers. This mechanism for ultrafast modification of the magnon frequencies comes directly from the electronic excitation, in contrast to indirect method which use the temperature dependence of the anisotropy field\cite{KJK02,HKK06}.

In conclusion, we have extended the domain of TDDFT simulations to include magnon dynamics in non-equilibrium systems. This opens the field of laser-coupled magnonics to \emph{ab-initio} theory. We first showed the prediction by TDDFT of element specific magnon modes with vastly different energies in Fe$_{50}$Ni$_{50}$ alloy. We then demonstrated three ways in which ultrafast laser pulses can control magnon dynamics: (1) selective destruction of particular magnon modes where the Ni or Fe modes could be selectively destroyed depending on the laser intensity, (2) laser driven destruction of magnon mode leading to transient non-collinear state of a ferro-magnet, and (3) OISTR-driven renormalization of the optical magnon frequency, where we found a linear dependence between the laser intensity (or moment transferred) and the decrease of the magnon frequency.\\ 

Due to computational restrictions, only 4 magnon modes were studied. However by elucidating the underlying physics of how these modes may be manipulated, we expect the observed effects are present throughout the BZ. Similarly, for the high wavevector modes studied, there is significant landau damping due to interaction with the Stoner continuum making experimental observation of these modes difficult.\\



In all cases the outcomes were achieved on ultrafast timescales thus demonstrating the potential of laser control of magnonics for future technology.  
In future work, we plan to study the more exotic magnons in AFM systems and magnetic insulators (which are more long-lived due to the lack of Landau damping) and high-wavevector modes excited via spin-transfer torque. \\

\section*{Methods}

The fundamental quantities in TDDFT are the density and the magnetization density which are defined as,
 
\begin{equation}
\begin{split}
n(\br,t) &= \sum_{i=1}^N |\phi_i(\br,t)|^2 \\
{\bf m}(\br,t) &= \sum_{i=1}^N \phi_i^* (\br,t) {\boldsymbol \sigma} \phi_i (\br,t)
\end{split}
\end{equation}
 
\noindent where ${\boldsymbol \sigma}$ are the Pauli matrices and $i$ is the joint index of {\bf k}-points and Kohn-Sham (KS) states. Within full non-collinear spin configuration, the KS orbitals, $\phi (\br,t)$ are treated as 2-component Pauli spinors propagated using the following equation:
    
\begin{equation}
\label{eq1}
\begin{split}
i \dfrac{\partial \phi_j (\br,t)}{\partial t} &= \Big[ \dfrac{1}{2} \Big( -i \bn + \dfrac{1}{c} \textbf{A}\ext (t)  \Big)^2 + v\s(\br,t) \\
&+ \dfrac{1}{2c} {\boldsymbol \sigma} \cdot \textbf{B}\s (\br,t) \Big] \phi_j (\br,t)
\end{split}
\end{equation}  


where $\textbf{A}\ext (t)$ is the vector potential representing the external laser pulse,  $v\s(\br,t) = v\ext(\br,t) + v\H(\br,t) + v\xc(\br,t)$ is effective KS potential consisting of external potential, $v\ext$, Hartree potential, $v\H$, and the exchange-correlation (XC) potential, $v\xc$. Additionally the KS magnetic field is ${\bf B}\s(\br,t) = {\bf B}\ext(t) + {\bf B}\xc(\br,t)$ the sum of external magnetic field plus laser magnetic field, ${\bf B}\ext$, and the XC magnetic field, ${\bf B}\xc$. 

To study magnons in real-time using TDDFT, we developed the method described in \cite{SEDGS20} where supercells commensurate with the wave-vectors, $\bq$, of particular magnon modes are constructed. In this work, we extend the method to include laser pulses, which can excite the electrons to a non-equilibrium state, allowing us to deduce the effect of this on the magnon modes. 


The magnon frequencies are obtained by Fourier transforming the transverse atomic moments following the laser pulse. The frequency of the unperturbed modes agrees with linear-response TDDFT calculations, thus validating the approach. While TDDFT is formally exact state-of-the-art method for treating magnonics in out-of-equilibrium systems, but the price to pay for such a treatment is that it is highly computationally demanding. This restricts the size of supercell that is practical and hence limits our study to high energy/high $\bq$ modes. However, the physics of the problem remains valid even at lower $\bq$ values. 

The $4$ atom super-cell of Fe$_{50}$Ni$_{50}$ is formed by extending the L$1_0$ primitive cell along the c-axis, where lattice parameters are $a=3.85 \text{\AA}$ and $c= 7.71 \text{\AA}$. The Brillouin zone was sampled on a {$\bf k$}-grid of $8\times8\times8$ and a time step of 1.209 attoseconds was used for time-propagating the orbitals using the algorithm presented in Ref. \onlinecite{DKSG16} . The adiabatic local spin density approximation to the XC functional was used. All simulations were done using the all-electron ELK electronic structure code\cite{elk}.

\section*{Data availability}
The ELK electronic structure code used for all calculations is freely available under the GPL license at \url{elk.sourceforge.net} . An installation guide and numerous examples are provided as part of the package. The \texttt{elk.in} input files are available upon reasonable request but can be easily constructed from the examples provided. 

\section*{Acknowledgements}

NS would like to thank SFB762 for funding. PE acknowledges funding from TRR227 (Project A04). The authors acknowledge the North-German Supercomputing Alliance (HLRN) for providing HPC resources that have contributed to the research results reported in this paper.


\bibliography{references} 

\end{document}